\documentclass[doublecol]{epl2} 

\usepackage{dcolumn}   
\usepackage{bm}        
\usepackage{amssymb, amsmath}   

\usepackage{graphicx}

\def \bs {\boldsymbol}

\newcommand{\abs}[1]{\lvert #1 \rvert}

\title{Detangling flat bands into Fano lattices}

\author{Sergej Flach\inst{1} \and Daniel Leykam\inst{2} \and Joshua D. Bodyfelt\inst{1} \and Peter Matthies\inst{3} \and Anton S. Desyatnikov\inst{2}}
\shortauthor{S. Flach \etal}

\institute{
\inst{1} New Zealand Institute for Advanced Study, Centre for Theoretical Chemistry \& Physics, Massey University,  0745 Auckland, New Zealand

\inst{2} Nonlinear Physics Centre, Research School of Physics and Engineering, The Australian National University, Canberra ACT 0200, Australia

\inst{3}Moscow Institute for Physics and Technology (State University), 9 Institutskiy pereulok, Dolgoprudny, Moskovskaya obl., 141700 Russia
}


\abstract{
Macroscopically degenerate flat bands (FB) in periodic lattices  host compact localized states which appear due to destructive interference and local symmetry. Interference provides a deep connection between the existence of flat band states (FBS) and the appearance of Fano resonances for wave propagation. 
We introduce generic transformations detangling FBS and dispersive states into lattices of Fano defects. Inverting the transformation, we generate a continuum of FB models. Our procedure allows us to systematically treat perturbations such as disorder and explain the emergence of energy-dependent localization length scaling in terms of Fano resonances.}

\begin{document}

\maketitle

{\sl Introduction ---}
The effect of interactions and disorder on wave transport in periodic potentials, such as electrons in crystals,  is strongly amplified if the bandwidth (kinetic energy) is small. A particularly interesting situation arises when some of the dispersion bands become strictly {\it flat} with macroscopically degenerate eigenstates. In this limit, any {\sl relevant} perturbation will lift the degeneracy and determine the emerging highly correlated and nontrivial eigenstates. A celebrated example is the fractional quantum Hall effect, which occurs as a result of the flat-band (FB) degeneracy of Landau levels of electrons in a magnetic field~\cite{FQH}. There is a growing effort~\cite{bergholtz13,parameswaran2013} to construct FB lattice models supporting new topological phases without the need of low temperature and external magnetic fields, which may be realized in diverse settings including ultracold atoms in optical lattices~\cite{bloch2008}, light propagation in waveguide arrays~\cite{christodoulides2003}, and exciton-polaritons in microcavities~\cite{masumoto2012}. These systems allow control over the interactions that successfully compete with the kinetic energy, and may lead to new wave-transport phenomena~\cite{wu2007, huber2010, vicencio2013}. Engineering FB lattice models has been extended to three-dimensional (3D)~\cite{goda2006}, 2D~\cite{bergman2008,huber2010,green2010}, and even 1D settings~\cite{richter,hyrkas13}.

A number of FB construction pathways using graph theory were suggested \cite{mielke,tasaki,richter}. 
They use compact states which are fully localized on several lattice sites~\cite{bergman2008,richter}. The origin of the compact flat-band states (FBS) is the destructive interference effectively decoupling FBS from the rest of the lattice, similar to the antisymmetric bound states embedded in and decoupled from the continuum in Ref.~\cite{bound} and geometric frustration in spin chains\cite{spin}. The interferometric nature of FBS suggests the appearance of Fano resonances~\cite{miroshnichenko10}, similar to the universal role of Fano interference in competition with bound states in the continuum~\cite{nanorib}, phase dislocations~\cite{vort}, and Anderson localization~\cite{FanoAnd}. The compactness of FBS significantly modifies disorder induced localization and metal-insulator transitions~\cite{goda2006,Nishino,chalker10} and may be instrumental in achieving topological Anderson insulators~\cite{AndTI}.

In this Letter we develop a generic detangling procedure of FBS from the dispersive part of the lattice, which allows to track the impact of perturbations in a systematic way. The number of unit cells involved in one irreducible FBS defines the FB class $U$ of the model. Here we transform and detangle the FBS and dispersive states into a lattice of Fano defects. Inverting the scheme, we derive a continuum of FB models for any FB class. In the case of an on-site disorder potential, the symmetric part of it lifts the FB degeneracy yet keeps the compact localization of FBS. The antisymmetric part yields Fano-induced Cauchy tails for the potential felt by the dispersive states. As a result, weak disorder enforces different energy dependent localization length scales, and highly nontrivial mode profiles at the FB energy. Scattering by perturbed FBS can be intuitively understood as a Fano resonance.

\begin{figure}
\includegraphics[width=\columnwidth, keepaspectratio, clip]{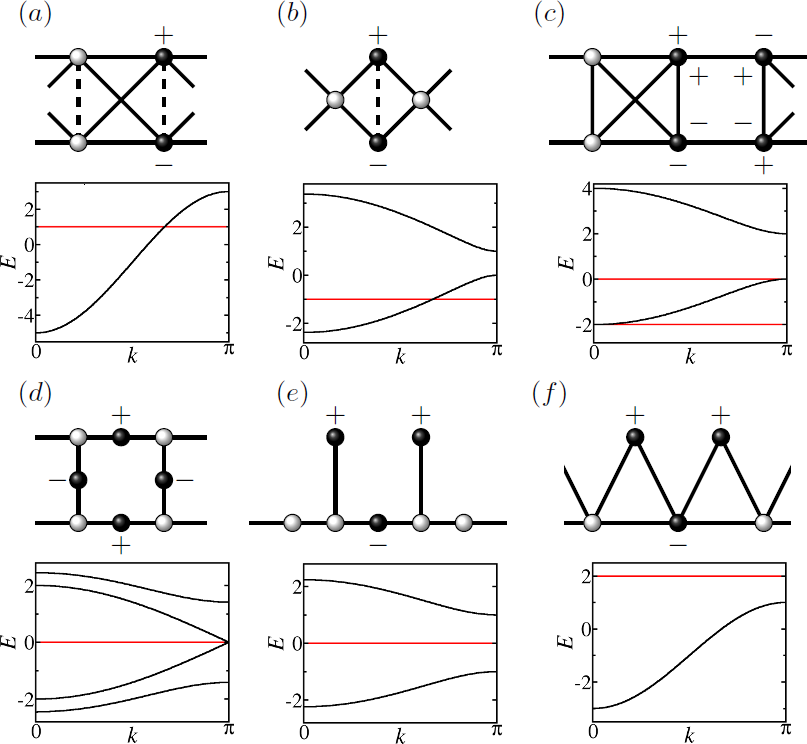}
\caption{(color online) 1D FB lattices. Circles denote lattice sites, solid lines are hopping elements of $t_{jj^{\prime}}$ with value 1, dashed lines are hoppings with tunable value $t$. Filled circles show the location of a compact localized state with identical wave amplitudes and alternating signs as indicated (all other lattice sites have strict zero amplitudes in such a FBS). The irreducible band structure is shown below each lattice. Onsite energies $\epsilon=0$, except in (f) where $\epsilon = 1$ for the upper row. Flat bands correspond to red horizontal lines. (a) cross-stitch $U=1$; (b) tunable diamond $U=1$; (c) 1D pyrochlore $U=1$; (d) 1D Lieb $U=2$; (e) stub $U=2$ \cite{hyrkas13}; (f) triangle $U=2$.}
\label{fig1}
\end{figure}

{\it Flat band models and compact localized states ---}
Consider a lattice wave eigenvalue problem of the type $E \Psi_j = \epsilon_j \Psi_j -\sum_{j'} t_{jj'} \Psi_{j'}$ where the wave components $\Psi_j$ are complex scalars allocated to points on a periodic lattice, the matrix $t_{jj'}$ defines some coupling between them, and $\epsilon_j$ are onsite energies. Such a generalized tight binding model produces a band structure for the eigenenergies $E_{\nu}({\bs k})$, $\nu=1,2,...,\mu$ (here ${\bs k}$ is a reciprocal Bloch vector, and $\nu$ counts the bands). Excluding the trivial case of just one band, $\mu=1$, we consider a model with at least one FB for which $E_{\nu}({\bs k}) = const$. Due to this macroscopic degeneracy, FB eigenvectors in the Bloch representation may be mixed to obtain highly localized FB eigenvectors~\cite{bergman2008,richter,vidal}. While there is no theorem which in general states that among all these combinations there will be compact localized eigenvectors, it is at least tempting to search for such cases~\cite{richter}. In fig.~\ref{fig1} we show that indeed for a set of known FB models, compact localized FB eigenvectors exist. We classify the compact localized FBS by the number $U$ of unit cells occupied by each state.

\begin{figure}[hbt]
\centering
\includegraphics[width=\columnwidth, keepaspectratio]{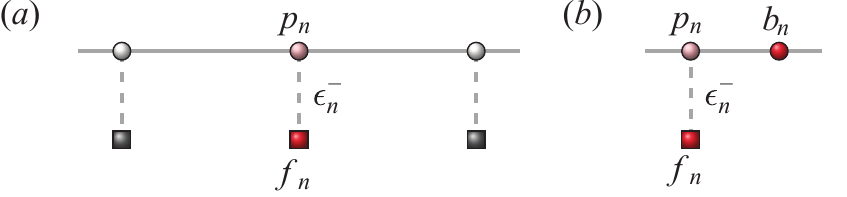}
\caption{(color online) Detangled Fano lattices. (a) Cross-stitch lattice from fig.~\ref{fig1}(a) detangled with eqs.~(\ref{eq:cross_trans_a}-\ref{eq:cross_trans_b}). Horizontal couplings are of strength $2$, and vertical couplings follow $\epsilon_n^-$. (b) Tunable diamond chain from fig.~\ref{fig1}(b).}
\label{fig2}
\end{figure}

{\it Detangling into Fano lattices ---}
The simplest 1D case with $\mu=2$ and class $U=1$ is the cross-stitch lattice, shown in fig.~\ref{fig1}(a). The amplitude equations read
\begin{eqnarray}
E \, a_n = \epsilon_n^a a_n - a_{n+1} - a_{n-1} - b_{n-1} - b_{n+1} - t \, b_n\;, \label{eq:cross_a}\\
E \, b_n = \epsilon_n^b b_n - a_{n+1} - a_{n-1} - b_{n-1} - b_{n+1} - t \, a_n \;. \label{eq:cross_b}
\end{eqnarray}
In the absence of a potential, $\epsilon_n^a=\epsilon_n^b=0$, there is exactly one flat and one dispersive band, whose relative positions are tuned with $t$:
\begin{equation}
E_{FB} = t, \quad E(k) = -4 \cos(k) - t \;. \label{eq:cross_disp}
\end{equation}
The flat and dispersive bands intersect if $|t|\le 2$. Introducing the transformation
\begin{eqnarray}
p_n = \dfrac{1}{\sqrt{2}}\left(a_n+b_n\right) , \qquad & f_n = \dfrac{1}{\sqrt{2}}\left(a_n-b_n\right) , \label{eq:cross_trans_a} \\
\epsilon_n^+ = \dfrac{1}{2}\left(\epsilon_n^a + \epsilon_n^b\right), \qquad & \epsilon_n^- = \dfrac{1}{2}\left(\epsilon_n^a - \epsilon_n^b\right), \label{eq:cross_trans_b}
\end{eqnarray}
we obtain a lattice with dispersive degrees of freedom $p_n$ and side-coupled Fano states $f_n$~\cite{miroshnichenko10},
\begin{eqnarray}
E \, p_n &=& \left(\epsilon_n^+ - t\right) p_n + \epsilon_n^- f_n - 2\left(p_{n+1} + p_{n-1}\right) \;, \label{eq:cross_fano_a}\\
E \, f_n &=& \left(\epsilon_n^+ + t\right) f_n + \epsilon_n^- p_n \;. \label{eq:cross_fano_b}
\end{eqnarray}
In the following, we refer to such lattices as ``Fano lattices'', see fig.~\ref{fig2}(a). Interestingly, such lattices with side-coupled defects also appear as models for charge transport in DNA \cite{dna0,dna}.

The transformation~\eqref{eq:cross_trans_a}-\eqref{eq:cross_trans_b} is a set of permuting local rotations, each in the $n$-th vector space $\{a_n,b_n\}$. If the potential $\epsilon_n$ satisfies the local symmetry $\epsilon_{n_0}^- = 0$, the corresponding Fano state $f_{n_0}$ decouples completely. If this symmetry is supported on all unit cells, $\epsilon_n^{-}=0$, then all Fano states decouple with individual energies $E_{fn}=(t+\epsilon_n^+)$. If, in addition, $\epsilon_n^+=\epsilon$ for all $n$, the Fano states form a FB.

{\it Generating FB lattices ---}
Let us invert the procedure. We choose a dispersive chain eq.~\eqref{eq:cross_fano_a} and set for simplicity $\epsilon_n=0$. We add a set of uncoupled Fano states $f_n$ with energies $E_{fn}$. We assign locally each $f_n$ to a site with $p_n$. We then perform local rotations (transformations) in the space $\{p_n,f_n\}$. 

Each rotation is parametrized by one angle $\theta_n$. For $\theta_n=\pi/4$ and $E_{fn}=-t$ we obtain the original cross-stitch lattice eqs.~(\ref{eq:cross_a},\ref{eq:cross_b}). Other values of $\theta_n$ generate modified cross-stitch lattices. An additional local potential $\epsilon_n^-$ results in purely local coupling of a state $f_n$ into the dispersive chain. If the energy $E_{fn}$ was in resonance with the dispersive chain, then the compact localized state $f_n$ will act similar to a Fano resonance in the Fano-Anderson model~\cite{miroshnichenko10}. If all Fano states $f_n$ are coupled into the dispersive chain, we obtain a Fano lattice.

\begin{figure}
\centering
\includegraphics[width=\columnwidth, keepaspectratio, clip]{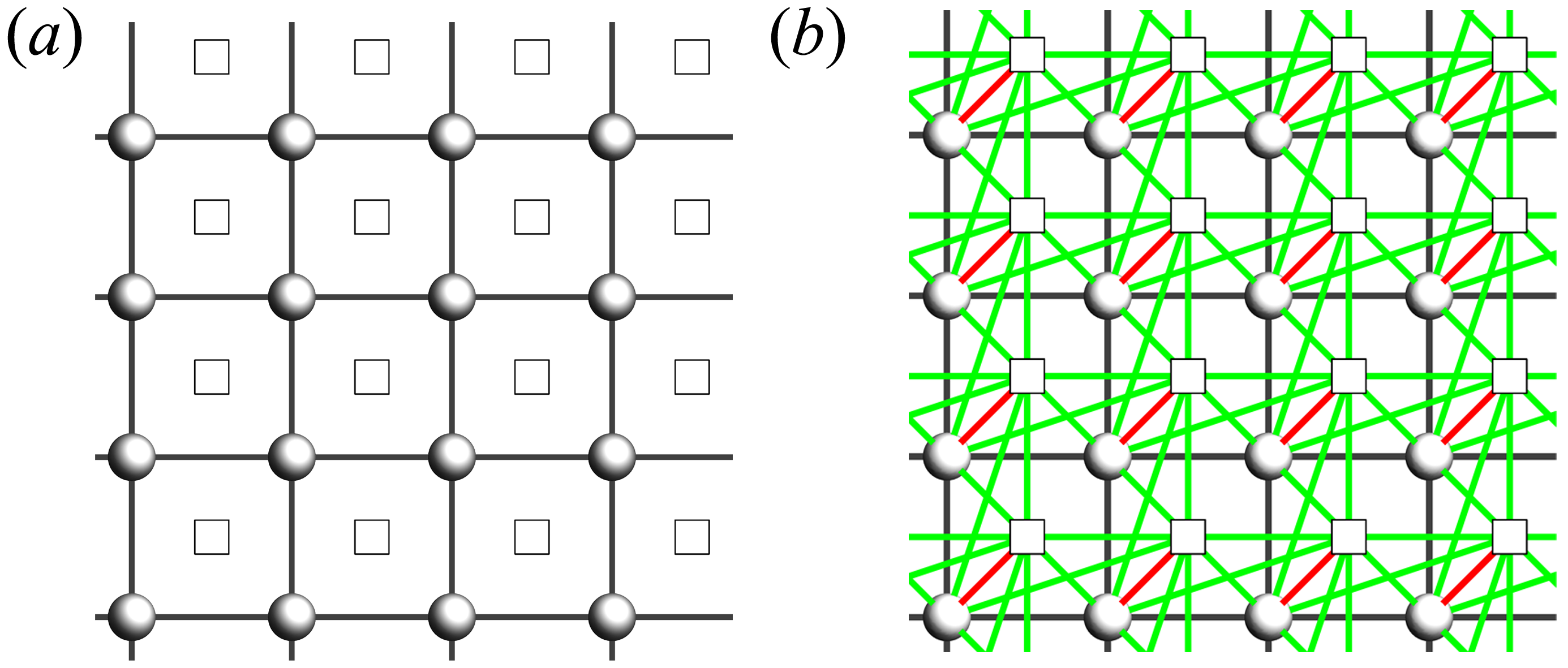}
\caption{(color online) (a) The irreducible Fano square lattice in two dimensions. (b) The rotated version. Bonds show only the connectivity, not the actual values. Newly appearing bonds are green and red. Red bonds indicate a tunable hopping strength which does not destroy the compactness of FBS.}
\label{fig3}
\end{figure}

Similar transformations can be performed with other models of class $U=1$ (fig.~\ref{fig1}(a-c)), and one example for the detangling of the diamond chain fig.~\ref{fig1}(b) is shown in fig.~\ref{fig2}(b). Moreover, we can generalize the construction procedure: consider any $d$-dimensional tight binding model with $m$ lattice sites per unit cell and $m$ dispersive bands. To each group of these $m$ lattice sites we assign $p$ Fano states, perhaps with different eigenenergies. Now we define a rotation in the corresponding $(m+p)$-dimensional vector space. If that is done in a translationally invariant way in all unit cells, we will obtain a complex looking $d$-dimensional lattice, which possesses $p$ flat bands. The graphical outcome of the simplest transformation for $d=2$ and $m=p=1$ is shown in fig.~\ref{fig3}. The dispersive lattice has energies $E(k_x,k_y)=-2(\cos k_x + \cos k_y)$. The FB energy can have any value.

If the Fano energies are nonuniform (e.g. random) along the lattice, or if the rotation angles are different for different unit cells, then the complex final lattice will even possess inhomogeneities. Nevertheless the underlying system remains translationally invariant in its dispersive part. That essentially concludes the $U=1$ case.

For $U\geq 2$ (e.g. fig.~\ref{fig1}(d-f)) the detangling procedure becomes hard, because the compact FBS do not form an orthogonal basis. Still, we can at least detangle in every $U$th unit cell in the 1D models in fig.~\ref{fig1} along the lines of the $U=1$ case, or in a similar way in two-dimensional models like the Lieb \cite{weeks12} or checkerboard \cite{chalker10} lattices.  With that we detangle $\frac{100}{U}$\% of the FBS, and will be left with the task of detangling the remaining fraction. However, if we are simply concerned with understanding the impact of disorder or similar perturbations on FB models, this partial detangling is already sufficient.

We can generalize our construction principle. Namely, we consider again a $d$-dimensional tight binding lattice with $m$ lattice sites per unit cell, and $m$ dispersive bands. We choose sets of $U$  (possibly neighbouring) unit cells, and assign $p$ Fano states to each. In the first procedure we rotate in the space of every $U$th assignment whose dimension is $Um+p$. Then we repeat the procedure, up to $U$ times. For example, for a 1D tight binding chain with $U=2$ and $p=1$ we assign in the first step a Fano state to two neighbouring sites (note that we have assigned in total $N/2$ Fano states, where $N$ is the number of lattice sites). Then we rotate in the subspaces of $2+1=3$ dimension each. In the second step we assign another $N/2$ Fano states in a similar manner, and rotate again. In general this produces a rather complex appearing $d$-dimensional lattice with many hoppings between nearest and next-to-nearest neighbors. 

{\sl Disorder, localization length, and Cauchy tails ---}
The detangling procedure and the Fano lattice representation allows us to systematically treat perturbations. Formally, the FB macroscopic degeneracy makes it hard to predict the impact of perturbations. In the detangled version, however, it can become rather easy and straightforward. An example is the case of onsite potentials $\epsilon_n$ which change the energy of each site of a lattice. Consider first the cross-stitch lattice fig.~\ref{fig1}(a) with $E_{FB}=t$. As shown above, for $\epsilon_n^-=0$ the Fano states remain decoupled, but their degeneracy is lifted since $\epsilon_n^+ \neq 0$. This can be generalized to any lattice with compact localized FBS. If the onsite energies are identical on all sites which involve a compact localized FBS, then the FBS stay compact and the Fano states are still decoupled. Therefore the local FBS structure dictates a certain local symmetry. The asymmetric potential part induces an interaction between the FBS and the dispersive states. In particular for symmetry-related uncorrelated random numbers $\epsilon_n^{a}=\epsilon_n^b = \epsilon_n$ with probability density distribution (PDF) $\mathcal{P}(\epsilon_n)=1/W$ for $|\epsilon_n| \leq W/2$ and $\mathcal{P}=0$ otherwise, the Fano states of the cross-stitch lattice stay decoupled, but acquire an energy spread of the order of $W$ around $E_{FB}$. At the same time the dispersive lattice (\ref{eq:cross_fano_a}) becomes Anderson localized with a localization length $\xi \sim 1/W^2$ for weak disorder $W \leq 4$ \cite{kramer93}. We remind that the localization length characterizes the spatial decay of an eigenstate, e.g. for the cross-stitch lattice $\Psi_{(a,b),n} \sim {\rm e}^{-|n|/\xi}$.

If now the symmetry constraint is relaxed, and $\epsilon_n^a$ is not anymore correlated with $\epsilon_n^b$ (but still all numbers have the PDF $\mathcal{P}$), then $\epsilon_n^- \neq 0$ and the Fano states are locally coupled into the dispersive chain. Due to the purely \emph{local} coupling, the Fano states can be eliminated and we obtain a new equation for the dispersive lattice:
\begin{equation}
\left[ E + t - \epsilon_n^+ - \frac{\left(\epsilon_n^-\right)^2}{E - t - \epsilon_n^+}\right] \, p_n = -2\left(p_{n-1} + p_{n+1}\right). \label{eq:7}
\end{equation}
If $\abs{E-E_{FB}} \le W/2$ (FB localization), the denominator in the LHS of eq.(\ref{eq:7}) produces heavy $1/z^2$ Cauchy tails. This happens because the PDF $\mathcal{W}$ of $z=1/\epsilon_n^+$ is
\begin{equation}
\mathcal{W}(z) =\frac{2}{z^2} \int \mathcal{P}\left (y\right) \mathcal{P}\left(\frac{2}{z} - y\right) dy \; .\label{cauchy}
\end{equation}
If $\abs{E-E_{FB}} \ge W/2$ (dispersive localization), Cauchy tails are absent, and the dispersive localization length  $\xi_{DB} \sim 1/W^2$. These two different energy windows will be present for any flat band at energy $E_{FB}$ in any $d$-dimensional FB lattice with additional diagonal disorder. For energies $\abs{E-E_{FB}} \le W/2$ the dispersive lattice part is dressed with Cauchy tailed disorder.

The FB localization length in the 1D case is then predicted to scale as $\xi_{FB} \sim 1/W $ for $|t| < 2$ when $E_{FB}$ is in resonance with the dispersive spectrum, $\xi_{FB} \sim 1/W^{1/2}$ for $|t| = 2$ when $E_{FB}$ is at the edge of the dispersive spectrum, and $\xi_{FB} \sim constant$ for $|t| > 2$ when $E_{FB}$ is in a gap outside the dispersive spectrum. The first two conclusions follow from previous calculations of the localization length scaling in pure 1D tight binding chains with onsite Cauchy disorder~\cite{lloyd1969,thouless1972,ishii73,altshuler00,titov03}.

In the gapped case, Fano states show a disorder in energy of the order of $W$, and an effective hybridization (hopping) between them of the order of $(\epsilon^-)^2 \sim W^2$, since one has to first excite a dispersive band state, and then return to the Fano states. That gives a vanishing localization length for $W \rightarrow 0$ according to the standard Anderson approach~\cite{kramer93}. However, at any finite $W$ one hybridization step always connects a Fano state to the dispersive band. Then the Fano state acts as a defect state with a detuned energy, and generates a corresponding exponentially localized state on the dispersive band states. These can back-couple into the Fano state system and generate the same exponential localization profile there as well. This third case therefore yields a localization length which does not depend on the strength of disorder $W$, but is entirely controlled by the detuning of the FB energy $E_{FB}$ away from the dispersive bands into the gaps of the spectrum. The localization length is then obtained simply from assuming a gapped defect state at energy $E_{FB}$ which is decaying into the dispersive lattice. For instance, for the cross-stitch lattice we obtain 
\begin{equation}
E_{FB} = -4 \cosh (1/\xi) - t \;. \label{gapxi}
\end{equation}

\begin{figure}
\centering
\includegraphics[width=\columnwidth, keepaspectratio]{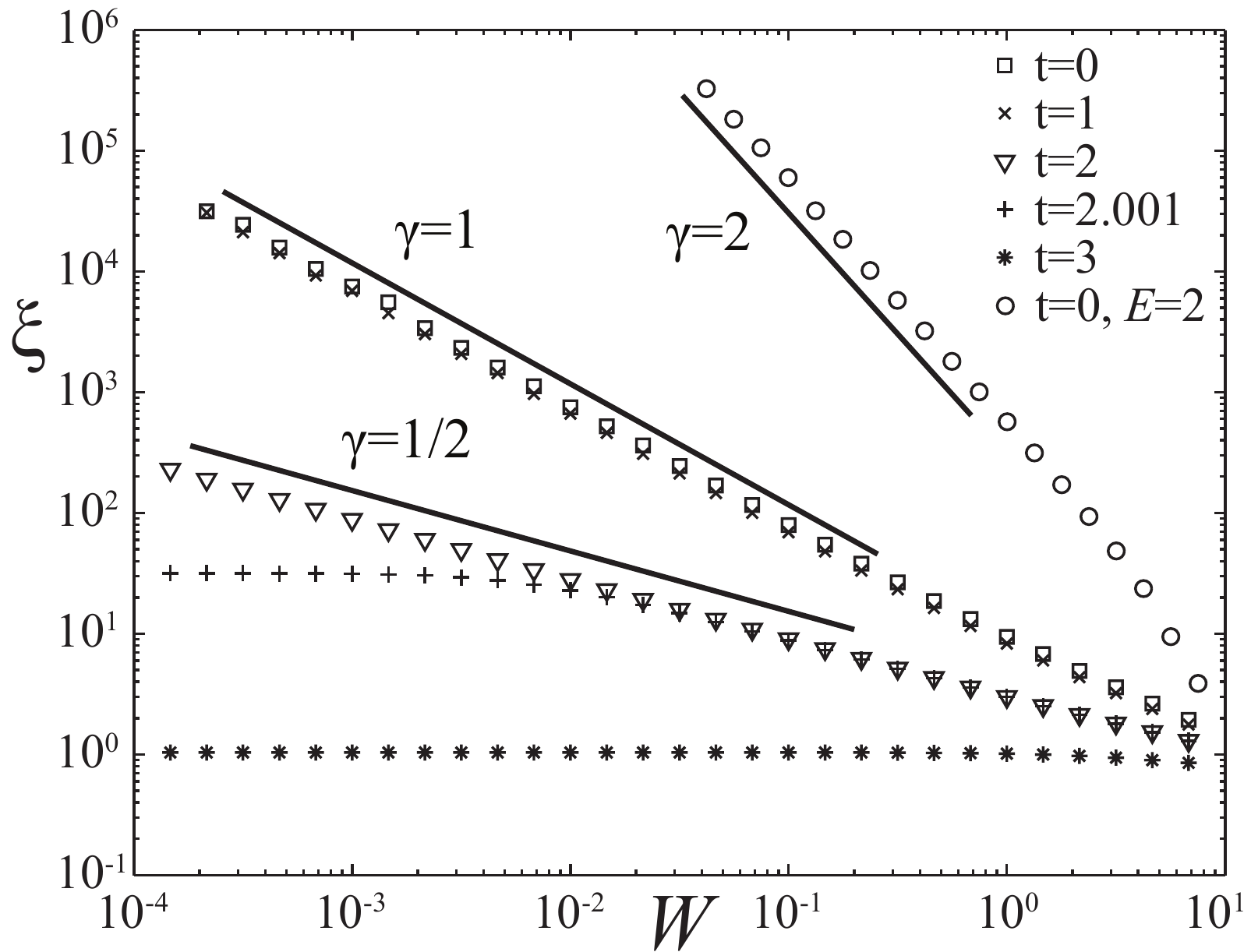}
\caption{Localization length scaling $\xi(W)\sim W^{-\gamma}$ for the cross-stitch lattice. For FBS in the continuum, $ \abs{E-t} < 2$, the scaling is $\gamma=1$ ($t=0,1$). Exactly at the continuum edge, $E=-t=-2$, the scaling is $\gamma = 1/2$ ($t=2$).  When the FBS are in the gap, saturation to constant values occurs ($t=2.001, t=3$). Note the transient following of the band edge law $\gamma=1/2$ down to $W\sim |t|-2$ from where on the gap location is resolved. Finally, for $t=0$ we show the dispersive localization length scaling at $E=2$, $\gamma=2$.}
\label{fig4}
\end{figure}
In fig.~\ref{fig4}  we show numerical computations of the localization length $\xi$ as a function of $W$ for the above cases of the cross-stitch lattice. We use standard transfer matrix methods by iterating a variant of eq.~(\ref{eq:7}) (see e.g. Ref.~\cite{kramer93}). We obtain excellent agreement with the predictions, observing the correct scaling laws. Moreover, in the gapped cases we obtain from eq.~(\ref{gapxi}) $\xi=31.6$ for $t=2.001$, and $\xi=1.04$ for $t=3$, in perfect agreement with the numerical results for small $W$. 

{\sl Sparse eigenstates ---}
Now we are in a position to discuss the shape of the disordered FB eigenstates. For that we have to consider the propagation of a wave at energy $E_{FB}$. While dispersing in the sublattice eq.~(\ref{eq:cross_fano_a}), the wave will encounter a Fano resonance with a FBS having an energy close enough to $E_{FB}$. This scattering event will involve a very strong population of the Fano state~\cite{miroshnichenko10}. Since the Fano state is coupled to the continuum with strength $W$ and the continuum has group velocities $\sim t$ (here $t=1$), the width of the Fano resonance is $W^2/t$. The Fano energies are distributed randomly in an interval of width $W$. We remind that Fano states appear at each unit cell in a Fano lattice. A given Fano state has then probability $W^2/tW=W/t$ to be in resonance with the propagation energy $E_{FB}$. If a localized state is characterized by a length $\xi$, we will count on average $W\xi/t$ Fano resonances in the volume $\xi$. Each of these resonances will contribute to a large peak in the eigenvector. The numbers of peaks in an eigenvector can be measured with the participation number $P=1/\sum_n(|a_n|^4+|b_n|^4)$. It follows that $P\sim W\xi/t$. If $E_{FB}$ is in resonance with the dispersive band, then $P$ becomes independent of $W$ in the limit of weak disorder, despite the fact that the localization length diverges as $\xi \sim 1/W$. Therefore disordered FB eigenstates have a sparse structure with a finite number of peaks and an increasing distance between them as the disorder weakens.
We test that for the cross-stitch lattice by computing the average over the participation number $P$ at the energy $E_{FB}$ for $t=0$ and different disorder strengths. We confirm that $P$ remains finite as $W \rightarrow 0$: $P(W=1)\approx 8$, and  $P(W=0.01)\approx 9$.

These results can be taken to higher dimensions $d$. A Fano state will still scatter in a similar way with its resonance width $W^2/t$ being independent of the dimensionality of the continuum. However an eigenstate will occupy now a volume of the order of $\xi^d$, yielding on average $W \xi^d / t$ resonances. In addition, the localization length is expected to diverge faster for weak disorder in $d=2$, and allow for mobility edges and complete divergence at finite disorder values in $d=3$.

For $d = 2$, even the conservative ansatz of $\xi \sim 1/W$ yields a divergence of the participation number of FB states at weak disorder, but nevertheless much slower than the growth of the localization volume $\xi^d$ itself. The localized eigenvectors will then have a growing localization volume, a growing number of peaks within, and a growing distance between these peaks, i.e. a growing sparsity, signalling a fractal structure of the FB eigenstates as discussed in \cite{chalker10}. Further, for a FB energy at the mobility edge in the $d=3$ case, not only fractal FB states, but perhaps a more intricate modification of the metal-insulator transition point can be expected. Thus, we expect qualitatively different behaviour compared to the inverse Anderson transition obtained in Ref.~\cite{goda2006}, where no dispersive bands were present.

{\sl Conclusions and outlook --- }
Previously the localization length at the FB energy $E_{FB}=0$ of the diamond chain fig.~\ref{fig1}(b) was evaluated~\cite{leykam13}. At this particle hole symmetric point, $\xi \sim 1/W^{1.3}$ was observed, distinct from the expected Cauchy law $1/W$. When $E_{FB} \neq 0$, away from the particle hole symmetric point, we found complete agreement with the Cauchy prediction. The singular behaviour at $E_{FB}=0$ remains to be explained.

We predict that in 2D and 3D, the impact of flat band disorder will be again the generation of heavy Cauchy tails in the effective disorder potential for dispersive waves. It will be therefore very useful to understand the impact of Cauchy tailed disorder in these dimensions.Furthermore, in these higher dimensions our procedure may be generalized to construct anisotropic Fano states and design lattices displaying direction-dependent localization\cite{rodriguez2012}.

Off-diagonal disorder can also be induced in the matrix elements $t_{jj^{\prime}}$, and can be even studied experimentally with microwaves propagating in networks of dielectric resonators~\cite{bellec13}. Similar to the onsite disorder, a locally symmetric off-diagonal disorder will not destroy the compactness of FBS, while asymmetric disorder will couple them back into the dispersive lattice, generating similar Cauchy tails and Fano resonances.

\acknowledgements

We appreciate useful discussions with O. Derzhko, A. Miroshnichenko, R. Moessner, and J. Richter. DL and ASD
acknowledge support from the Australian Research Council.

\end{document}